# POCKET CODE: A MOBILE APP FOR GAME JAMS TO FACILITATE CLASSROOM LEARNING THROUGH GAME CREATION

## B. SPIELER, A. PETRI, C. SCHINDLER, W. SLANY, ME. BELTRAN, H. BOULTON, E. GAETA, J. SMITH

## Introduction

Game jams are a way to create games under fast-paced conditions and certain constraints (Eberhardt, 2016; Deen, et al., 2014). The increase in game jam events all over the world, their engaging and creative nature, with the aim of sharing results among players can be seen in the high participation rate of such events (2013: 16,705 participants from 319 jam sites in 63 countries produced 3248 games) (Fowler, Khosmood and Arya, 2013). This promising concept can be easily transferred to a classroom setting.

Academic game jams are a kind of project work that fosters collaboration and at the same time results in understanding learning content from different subjects (Chandrasekaran, et al., 2012). Tools normally used in professional game jams (Suddaby, 2013), like the game engine Unity3d[1] or the computer graphics software blender[2] are either difficult to learn for young students or not available in schools. This paper argues that Pocket Code, a mobile app enabling one to program games within minutes, is easy to learn even for novices, and is applicable to different academic subjects. It seems to be a perfect tool for game jams. Children nowadays grow up with mobile devices, and feel comfortable using them. Considering the current prices and the forecast of the user penetration for smartphones in Austria, France, Germany, and the United Kingdom from 2014 to 2021 (Statista Market Analytics, 2016) as well as the difference in number of smartphone and tablet users in Western Europe in 2014 (eEmarketer;

---

[1] https://unity3d.com/

[2] https://www.blender.org/

February 2, 2015) and the current electronic device usage in Austria in 2016 (TNS Infraset, Google, 2016) one can conclude that smartphones will probably used more by students in future than the more expensive tablets or laptops. Further, a mobile app greatly facilitates research since relevant data can automatically recorded when uploading the games to the Pocket Code's code sharing web-platform (subsequently referred to as web-share).

This paper presents the general setting of a game jam, explains the practice of using Pocket Code in the school context, shows the aims of the project, and highlights the first experiments in performing Pocket Code Game Jams.

## Game Jams: An overview

Recent studies of game jams explored the collaborative nature (Chatham, et al., 2013) in combination with improvement in self-efficacy (Smith and Bowers, 2016), identified certain guidelines (Goddard, Byrne and Mueller, 2014), referred to game jam frameworks like Mechanics Dynamics Aesthetics (MDA) (Buttfield-Addison, Manning and Nugent, 2015), or investigated the motivation of jammers and their reasons for participation (Wearn and McDonald, 2016). Until now, less attention has been given to exploring game jams within an academic context e.g., students at the high-school level.

Goddard, Byrne and Mueller (2014) have identified several game jam characteristics, e.g., appropriate team size, where teams are formed (online or on-site), audience (professional or academic), timeframe (normally ranging from 24 to 48 hours (Moser, et al., 2014), occurrence (continuous or work hours), process (open, internal, or milestones), place (e.g., co-located), awards (for games or pace), constraints and submission (digital or presentations). The essential factor to frame a game jam is to define constraints for space and scope like a given theme or additional diversifiers, e.g., a local multi-player mode or to use materials found in the public domain. These diversifiers can provide small additional sub-goals to aim for (Global Game Jam®, 2016). All rules push participants to be fast, think creatively, work collaboratively, and finish a game within a given deadline (Kaitila, 2012).

Preston, et al. (2012) characterized a typical game jammer at the most popular game jam event: the worldwide Global Game Jam (Fowler, Khosmood and Arya, 2013), which plays a significant role in research. The participants are mostly male and already advanced in various areas (knowledge in at least one programming language or game-making software), with the motivation to meet potential business partners or to sharpen skills. This fact could lead to social pressure for a novice developer. Another point (Jaffa, 2016) about common game jams notes that jam participants need their own hardware and tools to create their projects and therefore puts participants without the ability to make their own tools at a disadvantage.

By contrast, in an academic setting, game jams allow students with common goals to work together while expressing individual ideas and creativity (Chatham, et al., 2013). Therefore, game jams cover various game-making disciplines, like programming, art, and design, and support learning by doing.

For game jams with an academic purpose the theme centres mostly on school topics (Goddard, Byrne and Mueller, 2014), where factors such as learning achievement, engagement, and persistence are important.

## Pocket Code: Creating games

Tools, like Scratch[3] or Snap[4] that were designed to help programming beginners through a visual programming language are already well known and adopted in computing classes all over the world (Meerbaum-Salant, Armoni and Ben-Ari, 2010). These visual programming languages keep the focus on the semantics of programming and eliminate the need to deal with syntactical problems. Pocket Code[5] is similar to Scratch but it can be directly programmed on the mobile device.

It is freely available at Google's Play Store and allows its users to create games, animations, music videos, and other kinds of apps on their smartphones. Pocket Code integrates with the device's sensors such as inclination, acceleration, loudness, or compass direction. It uses a visual, Lego®-style way to put code bricks together to form scripts.

For demonstrating the functionality of Pocket Code's concise user interface, a simple Pocket Code program is developed (see Figure 1). The program consists of two objects (left), the two looks used for animation (centre), and the bird's script that defines its behaviour (right).

---

[3] https://scratch.mit.edu/

[4] http://snap.berkeley.edu/

[5] http://www.catrobat.org/

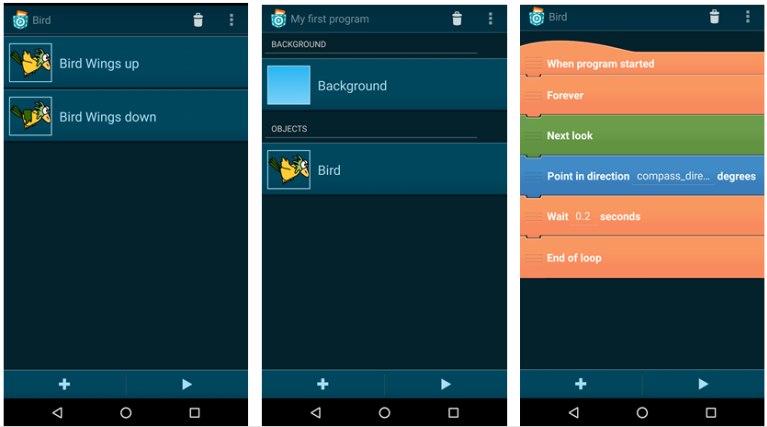

Figure 1.        Example Pocket Code program.

Every program has a number of objects and one background (which is a special object). Every object can hold a.) scripts to control the object, b.) looks, which can be changed and c.) sounds, to integrate music. The behaviour of the object and its looks and sounds can be controlled by scripts. The goal is to create a bird flapping its wings and always pointing to North, wherever the phone is pointing to.

This Pocket Code demo consists of two elements, a sky-blue background object and a bird object (Figure 1 (left)). The object bird has two different looks, which are used for the animation of its wings (Figure 1 (centre)). The bird's scripts section contains a single script that makes the bird flap its wings and updates the bird's direction to the North (Figure 1 (right)). The script consists of different-coloured bricks, indicating their originating brick category, e.g., control, motion, sound, looks, and data. The first brick, "When program started," is a trigger that starts the execution of the script whenever the Pocket Code program is started by the user. The "Forever" brick, with its delimiter "End of loop," represents an endless loop, meaning that every brick between "Forever" and "End of loop" is executed as long as the Pocket Code program is executed. The "Next look" switches the object's appearance from "wings up" to "wings down" and the "Point in direction " brick updates the object's direction. To use the compass direction, Pocket Code accesses the device sensor through the formula editor functionality. The last brick's purpose in the forever-loop is to slow down the animation rate. Therefore, a "Wait 0.2 seconds" brick is inserted, causing a 0.2s delay in this loop.

# The "No One Left Behind" project

Pocket Code has already been validated as an effective learning and teaching tool in the ongoing European project "No One Left Behind"[6] (NOLB), funded by the Horizon2020 program. During the feasibility study from September to December 2015, followed by the first cycle in Spring 2016, three pilot studies in Austria, Spain, and the UK were conducted. Each pilot targeted around 200 students between age 10-17, experiencing social exclusion problems. The Austrian study is dedicated to raise girls' interest in Science, Technology, Engineering, Arts, and Mathematics (STEAM)-related subjects and fosters social inclusion in class (Craig, Coldwell-Neilson and Beekhuyzen, 2013). In these ways, Pocket Code should enhance students' abilities across different academic subjects, and improve their computational proficiency, creativity and social skills.

The goals were measured in two ways:

1. Three quantitative surveys have been conducted: a pre-questionnaire before starting with Pocket Code, a questionnaire directly after the last Pocket Code unit, and a post-questionnaire about one month after the last Pocket Code unit. These surveys measured students' intention to use Pocket Code and possible barriers, differences in subgroups, gender and usability barriers related to Pocket Code. The results have been analysed via a descriptive content analysis and a user experience model.

2. The learning objectives defined by the teacher beforehand were measured against the learning outcomes. Games that have been uploaded to the web-share have been analysed towards learner achievement, collaboration, persistence, engagement, and amount of assistance/guidance needed. This data was collected through on-site observations, recorded by taking notes, videos, and photographs.

The results show that Pocket Code is easy to use (evaluation of the questionnaire) and fosters collaboration (most projects were done in groups of two), has the potential to help students' academic performance (105 out of 172 projects fulfilled the learning goal defined beforehand by the teacher), and, thus by acting as a supporting learning tool, leads to accomplishing academic curriculum objectives. Further formal assessment will be done during the 2nd cycle of the NOLB project starting in September 2016.

---

[6] http://no1leftbehind.eu/

Beside this research, Pocket Code is a convenient tool for game creation in game jams (Petri, et al, 2015) because of its general applicability. Pocket Code allows one to:

- create games within a short time span in fast paced and collaborative environments.
- merge programs among peers and transfer objects, code, looks, and sounds between projects via the "Backpack" functionality and therefore fosters distributed development.
- share the collection of programs through the upload to the web-share
- participate easily. According to a survey at the beginning of the study, 179 out of 187 pupils in Austria's pilot had their own mobile devices. For participation no costly hardware or tools are needed. This point facilitates the setup of Pocket Code game jams since there is only a minimal organization effort for teachers and schools.
- participate globally. Since Pocket Code is translated into 40 different languages, students all over the world can participate and use the app in their mother tongue. Students worldwide can be reached through online game jams and submission through the web-share.

For the NOLB project, the goals towards game jams include:

- Identifying benefits for students to run game jams in academic contexts.
- Identifying problems such as difficulties in generalizing results or missing functionality in Pocket Code.
- Holding several official online Pocket Code Game Jams to gain deeper understanding in setting up such events and discover potential obstacles.

## Game Jam Experiments and Results

The first two official Pocket Code Game Jams events were held during the European Code Week from 12th to 18th of October 2015 and during the International Computer Science Education Week from 7th to 13th of December 2015. The game jam aimed to engage female teenagers and introducing them to programming in a playful way (Ann and Comber, 2003). The theme for both jams was "Alice in Wonderland" because it seemed to fit for all genders and could be transferred to different subjects like Maths or literature. The game jams were conducted together with the MIT Scratch team. The first game jam was held using first year computer science students at Graz University of Technology. This event brought insights for the main event in December 2015 and was part of their homework. For the second game jam, 95 games were submitted (54.74% Scratch, 45.26% Pocket Code). Participants were told to fill out a questionnaire after submitting their games. (All games can be found at pocketcode.org with the hash tag #AliceGameJam[7]). Results show that 46.32% participants were female, and 44.21% had already some knowledge in programming languages like C++, Java or Python (13 participants), NXT programming (2 participants), in Scratch (3 participants) or Pocket Code (4 participants). The average age of the participants was 17 years. The 95 project submissions were created mostly at home (62.11%) or in schools (32.63%). Schools were encouraged to make "Alice in Wonderland" background knowledge available to students beforehand, and 75.79% of the participants mentioned in the survey that they liked the theme. The findings indicate that slightly more than half of the submissions (51.57%) were created in small teams (29.47% teams of 2; 4.21% teams of three; 17.89% teams that consisted of more than 3 team members), thus identifying the potential for enabling skills such as sharing, team problem solving, and cooperation. For these reasons, game jams in classrooms have the potential to support the development of the children's social and academic attitudes. Furthermore, their various talents are nurtured by building and enriching personal and collaborative knowledge, and becoming part of a wider community with different social and cultural perspectives (see submissions from different countries in Figure 2).

---

[7] https://share.catrob.at/pocketalice/search/%23AliceGameJam

| Country | Number of Submissions |
|---|---|
| Italy | 31 |
| India | 20 |
| Austria | 16 |
| Spain | 4 |
| United Kingdom | 8 |
| United States | 3 |
| Bosnia Herzegovina Canada, Egypt, Germany, Hungary, Philippines | 1 |
| didn't mention a home country | 17 |

Figure 2.    Number of submissions for the Alice Game Jam event per country.

Almost half of the participants spent 2 to 7 days working on their programs. (44.21%) and 29.47% spent only 2 to 5 hours on programming their games. This shows that the participants were willing to spend extra time (outside of the school) to program their games. Reasons why they participated in this game jam included (multiple answers were possible): "I liked the topic" (23), "I wanted to create a game" (32), "It was part of a school/university activity" (60), and "My friends participated" (7). Surprisingly, nobody chose that he or she wanted to develop their ability to code. Only two participants mentioned that there weren't satisfied with their outcome. The survey also showed that games were created across different school subjects like Maths, German or Chemistry (see example game in Figure 3). Therefore, game jams can be adapted to support learning and teaching strategies across different disciplines and obviously do not need be restricted to computer science classes.

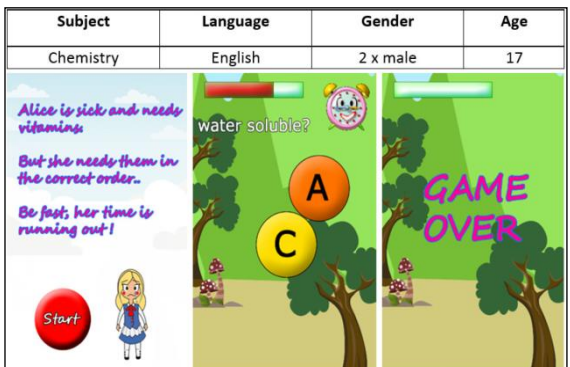

| Subject | Language | Gender | Age |
|---------|----------|--------|-----|
| Chemistry | English | 2 x male | 17 |

Figure 3.       Example project in Chemistry submitted for the Alice Game Jam event with the title "Sick Alice".

## Discussion and Conclusion

This paper argues that the programming app Pocket Code can support students in their learning goals and in combination with the promising concept of game jams for project works at schools. Further jams will need to be performed to provide a more precise matrix for recording the research. An upcoming feature of the Pocket Code app will include the integration of a customize software development kit (SDK) to track events within the app to define certain learning achievements like persistence through time spent in different parts of the app. Further, the paper shows that the concept of game jams works in a school context, but some additional challenges have been identified that must be addressed before the approach gains scientific relevance.

Challenges include that voluntary participation and intrinsic motivation, are also key factors of the play in game jams (Goddard, Byrne and Mueller, 2014). In a traditional school setting most teachers see a need for assessment and participation. To motivate more schools to participate in game jams further research is planned to design game jams especially for schools by providing helpful material, tutorials, and assets like graphics and sounds during the jam.

# Future Work

Plans include a coding-for-kids' roadshow over nine weeks on the main squares in cities throughout Austria, with morning and afternoon workshops on Pocket Code for school classes. The created games can be submitted for the Galaxy Game Jam[8] event in cooperation with Samsung, which will run from June until end of October 2016 (again during the European Code Week).

# Acknowledgements

This work has been partially funded by the EC H2020 Innovation Action No One Left Behind, Grant Agreement No. 645215.

---

[8] www.galaxygamejam.com

BibTex entry:

@InProceedings{SpielerICGBL2016,
author = {Spieler, B. and Petri, A. and Schindler, C. and Slany,
W. and Beltràn, M.E. and Bouldon, H.},
title = {Pocket Code: A Mobile App for Game Jams to facilitate
Classroom Learning through Game Creation}
booktitle = {In Proceedings of the 6th Irish Conference on
game-Based Learning},
series = {ICGBL '16},
isbn = {978-1535581455},
publisher = {CreateSpace Independent Publishing Platform},
location = {Dublin, Ireland},
month = {1-2 September, 2016},
year = {2016},
pages = {61-79}}